\documentclass[apjl,numberedappendix,letter]{emulateapj}
\usepackage[T1]{fontenc}
\usepackage{graphicx}

\makeatletter

\usepackage{ulem}

\usepackage{apjfonts}
\usepackage[T1]{fontenc}
\slugcomment{}
\shorttitle{}
\shortauthors{}

\makeatother

\def\degr{\hbox{$^\circ$}}
\def\radm2{rad~m$^{-2}$}

\begin{document}

\title{Rotation Measure Synthesis of Galactic Polarized Emission
with the DRAO 26-m Telescope}

\author{M. Wolleben\altaffilmark{1,5}}
\email{maik.wolleben@nrc.gc.ca}
\author{T. L. Landecker\altaffilmark{1}}
\author{G. J. Hovey\altaffilmark{1}}
\author{R. Messing\altaffilmark{1}}
\author{O. S. Davison\altaffilmark{1,2}}
\author{N. L. House\altaffilmark{1,3}}
\author{K. H. M. S. Somaratne\altaffilmark{1,4}}
\author{I. Tashev\altaffilmark{1,4}}

\altaffiltext{1}{National Research Council of Canada,
              Herzberg Institute of Astrophysics,
              Dominion Radio Astrophysical Observatory,
              P.O. Box 248, Penticton, British Columbia,
              V2A 6J9, Canada
}

\altaffiltext{2}{Department of Electrical and Computer Engineering,
              University of Alberta, Edmonton, Alberta, Canada}

\altaffiltext{3}{Department of Computer Science, University of Lethbridge, 
Lethbridge, Alberta, Canada}

\altaffiltext{4}{Department of Physics and Astronomy, University of British 
Columbia, Vancouver, British Columbia, Canada}

\altaffiltext{5}{Covington Fellow} 

\begin{abstract}

Radio polarimetry at decimetre wavelengths is the principal source of
information on the Galactic magnetic field. The diffuse polarized
emission is strongly influenced by Faraday rotation in the
magneto-ionic medium and rotation measure is the prime quantity of
interest, implying that all Stokes parameters must be measured over
wide frequency bands with many frequency channels. The DRAO 26-m
Telescope has been equipped with a wideband feed, a polarization
transducer to deliver both hands of circular polarization, and a
receiver, all operating from 1277 to 1762~MHz. Half-power beamwidth is
between 40 and 30 arcminutes. A digital FPGA spectrometer, based on
commercially available components, produces all Stokes parameters in
2048 frequency channels over a 485-MHz bandwidth. Signals are
digitized to 8 bits and a Fast Fourier Transform is applied to
each data stream. Stokes parameters are then generated in each
frequency channel.  This instrument is in use at DRAO for a Northern sky polarization survey. Observations consist of scans up and down the
Meridian at a drive rate of $\sim$0.9$^{\circ}$ per minute to give
complete coverage of the sky between declinations $-$30$^{\circ}$ and
90$^{\circ}$. This paper presents a complete description of the receiver and data acquisition system. Only a small fraction of the
frequency band of operation is allocated for radio astronomy, and
about 20\% of the data are lost to interference. The first 8\% of data from the survey are used for a proof-of-concept study, which has led to the first application of Rotation Measure Synthesis to the diffuse Galactic emission obtained with a single-antenna telescope. We find rotation measure values for the diffuse emission as high as $\sim \pm100$ \radm2, much higher than recorded in earlier work. 
\end{abstract}


\section{Introduction}

All extensive surveys of the Galactic polarized emission to date have
measured polarization properties of the emission at one or a few
frequencies over narrow bandwidths. Recent examples at decimetre
wavelengths are the DRAO Low-Resolution Polarization Survey
\citep{2006A&A...448..411W}, the Argentinian polarization survey
\citep{2008A&A...484..733T}, and the Effelsberg Medium Latitude Survey
\citep{2004mim..proc...45R}.  However, it is widely accepted that the
appearance of the polarized sky at these wavelengths depends less
on the signal properties at the point of emission and more on Faraday
rotation along the propagation path; that appears to be true for the
Milky Way emission at all frequencies up to about 5~GHz. If
polarization surveys do not contain information on rotation measure
(RM), the quantity that is significant for physical understanding of
the interstellar medium (ISM), their value is limited.

Recent years have seen the development and application of wide-band
feeds and receiver systems. Advances in signal-processing technology
have led to construction of digital full-polarization correlators, 
often based on Field-Programmable Gate Arrays (FPGAs) or
similar devices. In combination with commercial analog-to-digital
converters (ADCs) with high sampling rates, often exceeding 1~GHz,
FPGAs can be used to realize digital polarimeters with large input
bandwidths and high dynamic range. Wideband spectro-polarimetric data
can now be easily acquired.

Methods for the analysis of such data were
articulated over four decades ago by \citet{1966MNRAS.133...67B} but
the technical limitations of the time made the observations and their
analysis extremely difficult. \citet{2005A&A...441.1217B} have recently
developed Rotation Measure Synthesis (RM-Synthesis), and it has been shown to be a
powerful tool for the analysis of wide-band polarization data 
\citep{2005A&A...441..931D,2007A&A...461..963S,2007A&A...471L..21S,
2009A&A...494..611S,2009arXiv0904.0404B}. The application of this technique requires correctly calibrated, multi-frequency polarization data and has to date been exclusively applied to observations obtained with synthesis telescopes.

The output from multi-frequency polarization observations is a data cube of images at many wavelengths, $\lambda$. Application of RM-Synthesis transforms this data cube into an RM-Synthesis cube. The first and second dimensions of an RM-Synthesis cube are the coordinates on the sky, which are R.A. and DEC for the data here presented. The third dimension is Faraday depth, $\phi$ [\radm2], which is often used in RM-Synthesis to replace RM:
\begin{equation}
 \phi = 0.81 \int_0^d B_\parallel\,n_e\,\mathrm{d}l,  
\end{equation}
where $B_\parallel\,[\mu\mathrm{G}]$ is the magnetic field component parallel to the line-of-sight, $n_e\,[\mathrm{cm}^{-3}]$ the electron density, and $l\,[\mathrm{pc}]$ the distance along the line-of-sight. Choosing a value for $\phi$ the observed polarization vector in each frequency channel is rotated by ${\phi}{\lambda^2}$ and the rotated vectors are coherently added to obtain an image of polarized intensity at that value of Faraday depth. It is important to note that Faraday depth is usually not related to physical distance.

Significant wideband spectro-polarimetric studies have been made with
aperture-synthesis telescopes (see references above), but absence of information on 
zero-levels in Stokes U
and Q causes non-linear effects in polarization angle and polarized
intensity, making interpretation always difficult, and sometimes
impossible. Incorrect zero levels can turn polarized objects into
apparently unpolarized ones, and can render RM values for the
diffuse Galactic emission meaningless. If the region mapped is of
significant extent aperture-synthesis observations must be
complemented with absolutely calibrated single-antenna data to provide
correct zero levels and information on the largest structures.

Taking advantage of the technical developments described above, a new
generation of polarization surveys with single antennas is underway,
bringing the field into the era of 3-D polarimetry. Multi-frequency
polarization surveys with single-antenna telescopes include GALFACTS
with the Arecibo telescope from 1225 to 1525~MHz
\citep[][]{2005AAS...20719207G}, STAPS with the Parkes telescope from
1296 to 1804 MHz (PI M. Haverkorn), and S-PASS also with the Parkes
telescope from 2188 to 2412 MHz \citep{2008arXiv0806.0572C}. 

The work described here is part of a new initiative, the Global
Magneto-Ionic Medium Survey (GMIMS), a major project to survey the
entire sky from the northern and southern hemispheres covering 300~MHz
to 1.8~GHz with many thousands of frequency
channels using large single antennas \citep{2008arXiv0812.2450W}.  The focus of this paper is on
the technical realization of spectro-polarimetry using the 26-m
Telescope at the Dominion Radio Astrophysical Observatory (DRAO). The
telescope has been equipped with a wide-band receiver and an
FPGA-based FFT polarimeter. The data gathered to date have led to the
first application of RM-synthesis to single-antenna polarization data,
and we present those results and give a general interpretation.  The data
presented form the proof-of-concept study for the survey now in progress.

\section{Telescope, feed and receiver}

\subsection{Reflector and feed}

The axi-symmetric reflector, of diameter 25.6-m, is equatorialy mounted.The mesh surface has an accuracy of $\sim$0.5~cm rms. The receiver is mounted at the prime focus on a fibreglass tripod. The telescope can observe the sky between $-30^{\circ}$ and $+90^{\circ}$ in an almost fully automated fashion.  

The feed is scaled from the design of \citet{1972ElL.....8..474W}, which was developed for the Effelsberg 100-m Telescope. The DRAO 26-m Telescope has the same ratio of focal length to diameter, ${f/d}={0.3}$, and the feed provides excellent performance at its design frequency, 1420~MHz, where the aperture efficiency achieved is $\sim 0.53$ and the beam is circular with half-power width of $36'$.

Feed performance has been computed across the operating band \citep{Smegal}. Feed radiation patterns are highly circular (with equal half-power widths in the $E$- and $H$-planes) at 1400 MHz.  At lower frequencies the $H$-plane pattern is wider than the $E$, and conversely for frequencies above 1400~MHz. This leads to increased instrumental polarization at frequencies away from 1400~MHz. Figure 1 shows the beamwidth of the telescope across the entire frequency band. The small variations in beamwidth are real (they are highly repeatable) but their cause is unknown.

\subsection{Polarizer}

\begin{figure}[tbp]
  \includegraphics[clip, width=\columnwidth, bb=34 50 710
  525]{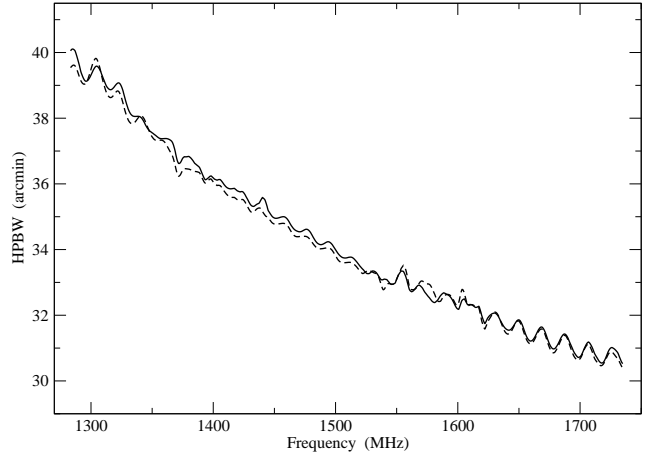}
  \caption{The half-power beamwidth (HPBW) of the DRAO 26-m telescope for RR (continuous) and LL
  (dashed) correlation products determined by fitting Gaussian brightness distributions to a few observations  of our primary and secondary calibrators. }
\label{figure1}
\end{figure}

The receiving system is designed to accept all the power from an incoming signal and to produce outputs that describe its polarization state. Two antennas of orthogonal polarization, whether linearly or circularly polarized, will together collect all the incoming power. Given the {\it{a priori}} knowledge that the linearly polarized component of synchrotron emission dominates any circularly polarized content \citep{1970ranp.book.....P} we designed the receiving system to measure small linearly polarized components in largely unpolarized signals. The two channels of the receiver accept right-hand circular polarization (R) and left-hand circular polarization (L). Stokes parameters $Q$ and $U$ that describe linear polarization can then be derived by correlation techniques.

Consider the incoming signals as the sum of two circularly polarized field components

$${E_r} = {{E_R}{\thinspace}e^{{\thinspace}i{\omega}t}},
{\thinspace}{\thinspace}{\rm{and}} \eqno{(2)}$$

$${E_l} = {{E_L}{\thinspace}e^{{\thinspace}i{\omega}t + {\delta}}}. \eqno{(3)}$$

In terms of these components the desired outputs from the polarimeter are

$${I} = \frac{1}{2}{({{{E_L}^2} + {{E_R}^2}})}, \eqno{(4)}$$

$${Q} = {{E_L}{E_R} {\rm{cos}}{\delta}}, {\thinspace}{\thinspace}{\rm{and}} \eqno{(5)}$$ 

$${U} = {{E_L}{E_R} {\rm{sin}}{\delta}}. \eqno{(6)}$$
${V} = {{\frac{1}{2}}({{{E_L}^2} - {{E_R}^2}})}$ is small and we will ignore it.

The alternative scheme, receiving two linear polarizations, leads to the derivation of $Q$ as the (small) difference between two large quantities, and is therefore quite susceptible to instrumental effects such as gain variations between receiver channels.

The operation of the feed as a collector of circularly polarized signals can be more easily understood by considering it as a transmitter and showing that it radiates circularly polarized signals. An orthomode transducer combines two monochromatic signals A and B of equal amplitude from coaxial inputs, launches them as separate signals in rectangular waveguide and then combines them in a single square waveguide where they travel in orthogonal (non-interacting) TE$_{10}$ and TE$_{01}$ modes. A power splitter in square waveguide converts A and B into $\frac{1}{2}(A+B)$ and $\frac{1}{2}(A-B)$. A differential phase shifter, also built in square waveguide, provides a $90^{\circ}$ relative phase shift between these signals and so generates outputs $\frac{1}{2}(A+iB)$ and $\frac{1}{2}(B-iA)$, which are circularly polarized components. A smooth transition from square to circular waveguide transfers these two outputs to the feed, from which they will emerge as circularly polarized signals in free space.

By reciprocity this device will collect the two circularly polarized components $E_l$ and $E_r$ described in equations 2 and 3 from an incoming wave and will deliver them to the coaxial inputs of the two receiver channels. In the following we denote $E_l$ simply as L and $E_r$ as R, and recognize that they are broadband signals. Powers will be denoted as: $RR=E_r\,E_r^\ast$, $LL=E_l\,E_l^\ast$, $RL=Re(E_r\,E_l^\ast)$, and $LR=Im(E_r\,E_l^\ast)$.

The return loss of the polarizer at the receiver inputs is 20 dB: only 1\% of the power in a signal injected into the polarizer at either port returns to that port. The isolation between the two ports is better than 30 dB (0.1\% of signal power). The dissipative loss is less than 2\% (0.1 dB) and the axial ratio of the device (measured by transmitting from it and analysing the radiated signal) is of the order of $\pm$10\%.

\subsubsection{Phase Shifter, Power Splitter, and Orthomode Transducer}

The phase shifter is built in square waveguide ($12.7 \times 12.7$~cm); its overall length is 
$\sim 60$~cm. Dielectric slabs line all four walls, their dimensions 
chosen to give a difference in insertion phase as close as possible to 
90$^{\circ}$ between orthogonal TE$_{10}$ and TE$_{01}$ modes in the 
waveguide. The design (developed with electromagnetic simulation software) 
gives a phase difference of $90^{\circ} \pm 2^{\circ}$ and in practice 
a performance of $90^{\circ} \pm 5^{\circ}$ is achieved. This design is based on the device developed by
\citet{1997IEEE...7..150S} using  corrugations in the waveguide walls  to provide the differential phase
shifts across the band 18.9 to 26.5~GHz. The fractional bandwidth of the two devices is similar
(essentially an entire waveguide band) and the performance is
similar. In our case use of corrugations would have resulted in lower
loss but the overall length would have been about three times
larger. The dielectric that we used was Teflon (dielectric constant
${\epsilon} \approx {2.3}$).  We explored designs using materials with
higher dielectric constant in the hope of reducing the length of the
device, but were unable to make an improvement, either in size or
performance, over the design using Teflon.

Dissipative loss of the phase shifter is slightly unequal in the two orthogonal modes because
more dielectric is used in one plane than in the other. The losses are
about 0.05 and 0.1~dB.


The input and output of the power splitter are both in square
waveguide, but the orientations differ by 45$^{\circ}$. The device is
fabricated from eight flat triangular plates bolted together. Its
length is $20.6$~cm.

The orthomode transducer is a commercially available device, originally
developed for the Very Large Array by Atlantic Microwave (their model number
OM6500). It has a square waveguide input and two outputs in rectangular 
WR650 waveguide.  Fabricated from cast aluminum, it is a low-loss device (loss is probably less than 0.05 dB or 1\%).

\subsection{Receiver}

\begin{figure}[tbp]
  \includegraphics[clip,width=\columnwidth]{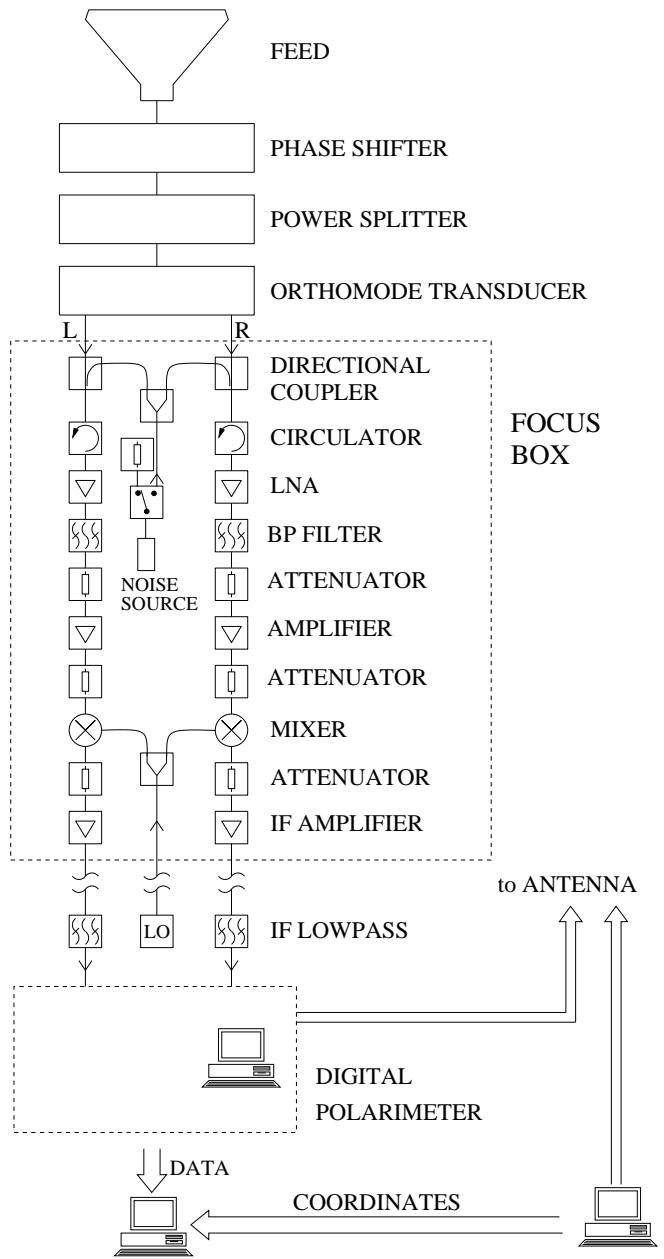}
  \caption{Block diagram of the receiver and signal-processing system.
  Parallel receiver paths process L and R signals. LNA denotes
  low-noise amplifier and CAL denotes calibration noise source.}
\label{figure2}
\end{figure}

The receiver is mounted at the prime focus in an insulated, temperature-controlled box. The air temperature inside the focus box is kept at $25.0\pm 0.1$ ${^\circ}$C. The two-channel receiver and signal processing system are depicted
schematically in Fig.~\ref{figure2}. Directional couplers permit
injection of noise calibration signals. Circulators (unidirectional devices which allow the low-noise amplifiers
to receive signals from the antenna ports but isolate the amplifiers from
reflections from the polarizer and feed) preceding the
first amplifiers prevent backscattering and thus reduce cross-coupling between the two hands of polarization, which would lead to instrumental polarization. The low-noise amplifiers (LNAs) are commercial uncooled
HEMT amplifiers with noise temperature ${\sim}$35~K, but elements
preceding them add about 30~K to their noise temperature.  The
receiver follows a conventional design, with a single-stage
down-conversion taking the RF band of 1277 to 1762~MHz to
baseband. The 3-dB bandwidth of the system is 485 MHz. The overall system temperature, including receiver noise, ground and atmospheric emission is estimated to be $\sim 140$~K.

\section{The FPGA-Based Polarimeter}

\begin{figure}[tbp]
  \includegraphics[clip, width=\columnwidth]{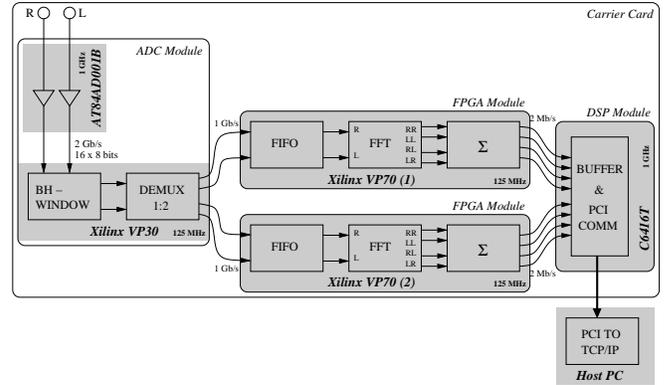}
  \caption{A block diagram of the polarimeter. See text.}
\label{figure3}
\end{figure}

To save development time and minimize engineering effort the polarimeter was implemented using off-the-shelf FPGA and ADC cards from Sundance Multiprocessor Technology Ltd.\footnote{The system is comprised of the following components: the carrier card (SMT310Q), two VP70 FPGA modules (SMT398-VP70-6), a DSP module (SMT395-VP30-6), and an ADC module (SMT391) piggy-back on another FPGA module (SMT338-VP30-6). For more detailed information we refer to Sundance documentation or {\url www.sundance.com}} (UK). This approach provided an integrated hardware and software solution that was relatively low cost and supported on a standard PC running Windows XP. The specifications for the polarimeter are given in Table~\ref{table1}. As shown in Figure~\ref{figure3} the hardware consists of a carrier card, an ADC module, and two FPGA modules. The ADC module has one Atmel ADC and a Xilinx VP30 FPGA. The ADC has two 8-bit converters and samples at 1 Gsps. It outputs its data at 125 MHz as two 8-bytes fields. These are received by the VP30 which applies weights to the input samples in the time domain (windowing) and outputs a frame of contiguous samples in time. These frames are distributed equally between the two FPGA modules on the carrier board, half going to one FPGA and half to the other. Each FPGA module has a Xilinx VP70, which first buffers the ADC fields into a frame in a FIFO (first-in first-out shift register). The data are clocked at a rate of 125 MHz. Fast Fourier Transform is perfomed on frames, which are then summed over an integration period. The integrated power spectra are sent to the Host PC over a 33 MHz PCI bus. The FPGA firmware is programmed in VHDL, and synthesized and compiled using Xilinx ISE. The processing firmware is divided into separate modules and these communicate in a standard way using FIFOs. This approach permits a high-level firmware management and inter-module communication software tool, called Diamond from 3L\footnote{{\url www.3l.com}} (UK), to be used.

\begin{deluxetable}{lr}[tbhp]
\tablecaption{Hardware specifications of the FPGA-based polarimeter}
\startdata
\hline
ADC sampling rate & max 2 x 1 GS/s \\
ADC resolution & 8 bit\\
Input bandwidth & 0 - 500 MHz \\
FFT Core & Xilinx FFT v4.0 \\
Number of channels (FFT) & 2 x 4096 (complex) \\
Number of channels (Output) & 4 x 2048 (real)\\
Precision (internal) & 2's complement, 54 bit \\
Precision (output) & 2's complement, 32 bit \\
Minimum integration period & 25 ms \\ 
Telescope interface requirements & noise source control 
\enddata
\label{table1}
\end{deluxetable}

\begin{figure}[tbp]
  \includegraphics[clip, width=\columnwidth]{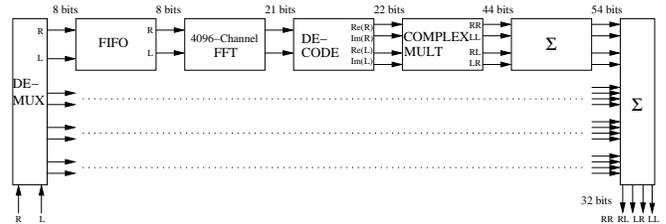}
  \caption{A block diagram of the FFT code. See text.}
\label{figure4}
\end{figure}

A block diagram of the polarimeter digital signal processing is shown in Figure~\ref{figure4}. The processing consists of windowing and Fast Fourier Transforming the two polarization streams, and then cross-multiplying and integrating these to produce the four polarization products (RR, LL, RL, and LR) as power spectra. A windowing function is applied to the data to minimize leakage from one channel into adjacent spectral channels. This is especially important for removing strong narrow band RFI. While a rectangular window yields the best frequency resolution and highest signal-to-noise ratio, spectral leakage into nearby channels is high, about -30 dB. Strong RFI in one channel can easily spread into many adjacent channels and swamp the sky signal. In the current design the Blackman-Harris (BH) window-function is used for weighting samples in the time domain. This window function yields a spectral leakage of less than -60 dB. This effectively suppresses leakage and baseline fluctuations due to narrowband RFI. It is implemented as a lookup table in the VP30 on the ADC module. One disadvantage is that it discards 50\% of the data and lowers the signal-to-noise ratio accordingly. In the future we will replace the window with polyphase filter techniques (e.g. WOLA) as described by, e.g., \citet{2006SPIE.6275E..33K}.

A 4096-point complex FFT is used to transform the windowed ADC data into spectra. This function is provided in the Xilinx supplied function library and is highly optimized. Because the VP70's are clocked at 125 MHz, eight FFT's are needed to process a 1 GHz data stream of R and L polarized data. Four FFT are performed on each VP70 FPGA. In this scheme an FFT block receives 1/8 of the frames from the ADC FPGA. As well, since the R and L polarization data are real, one 2048-point complex FFT can be used to compute the 2048-point FFT's for each polarization. The technique of using a complex FFT to compute two real FFTs is well known \citep[e.g.][]{nussbaumer} and involves placing 2048 points of R polarization data in the real part of the FFT and 2048 points of L polarization data in the imaginary part of the FFT. After the FFT is complete the results for each polarization are recovered yielding two 2048-point complex spectrums one for the R and L polarizations, respectively.

The complex spectra are then multiplied and summed to yield the integrated 2048-point power spectra for the four polarization products, namely RR, LL, RL, and LR. In the current implementation sums of 760 spectra form integrations with a period of 25 ms. The integrated spectra, comprising 8192 signed four-byte integers, are sent to the host PC over the PCI bus. The average data rate is about 1.3 MBps which is easily handled on a 33MHz PCI bus. 

\section{Electronic Gain and Phase correction}
\label{egainphase}
 
Amplifiers suffer from gain and phase variations which cause the
electronic gains to vary. Phase shifts of one hand of polarization relative to the other can
also be introduced by variations in cable length due to
temperature changes. These gain and phase drifts are
corrected using a calibration signal, a noise signal injected into the signal path ahead of the first amplifier through a directional coupler. An RF switch
between the noise source and the coupler is used to switch the
calibration signal on and off, synchronized with the polarimeter. A switching time of 200~$\mu$s is is allowed for between integrations. An integration packet
of 50~ms length consists of two integrations with cal-on and cal-off.

The actual intensity of the calibration signal for an integration
packet is determined by calculating the difference between the cal-on
and the corresponding cal-off spectrum. The equivalent noise
temperature of the calibration signal is approximately 20~K. The rms
noise in a single channel and a 50~ms integration is about 8\% of that
and is much larger than the expected gain
fluctuations. Therefore, the average over 200 channels and 90s (45s of
the preceding and 45s of the following integrations) of the
calibration signal is taken, which reduces the noise in the
calibration signal temperature to an acceptable value of 3 mK.

The calibration signal is split and equal-amplitude components are injected into the L and R paths. This is equivalent to a signal with constant polarization angle. The
sky values are then expressed relative to the measured intensity and
polarization angle of the calibration signal.

\section{RFI Mitigation}

Only a small fraction of our observing band lies within protected allocations for radio astronomy. Hence, RFI mitigation becomes a major, if not a limiting, factor in our observations. RFI is present over a wide range of intensities, from very strong satellite emissions (GPS and GLONASS among others), to extremely weak RFI which is seen only in the final data processing steps after sufficient averaging. The same can be said about the time-domain: in some channels RFI can be present at all times, while it is sporadic in others. In order to minimize the level of locally generated RFI the observatory buildings are screened, which attenuates radio signals in L-band by about 
25~dB. Known sources of RFI at the Observatory, such as computers, printers, laboratory equipment, and digital signal-processing equipment are enclosed in shielded boxes or are operated inside Faraday-cage screened rooms to further suppress their emission (by at least 50 dB). 

We found that RFI is more severe at low elevations when the telescope points south than at the Zenith. This may be due to two effects: geostationary satellites lie at low elevation as seen from Canada, and the spillover sidelobes look at the observatory building and pick up terrestrial RFI, most likely produced by the surrounding community. In contrast, with the telescope at the Zenith the spillover lobes see the ground. Rather than to try to correct for RFI in any way, our approach is to flag any RFI found in the data and use the remaining (interference-free) part of the spectrum for data analysis. We applied a three-stage RFI search to the data at different places in the data processing chain, reflecting the increase in sensitivity due to averaging. The first two stages are part of the real-time processing chain, and the last stage is applied to the final maps. If RFI is detected in one or more correlation products, all four correlation products are flagged for that channel. 

The first algorithm detects RFI in the time domain. It detects strong, time-variable signals and flags affected channels until 60 seconds after the last detection. On average about 10 to 20 channels are flagged in this process. A high threshold at this stage is necessary to avoid erroneous flagging of good data (for example, while scanning across a bright point source which raises the baseline rapidly). The second algorithm works in the frequency domain. A simple median filter is used to detect signals above or below a certain threshold. Naturally this algorithm is more successful than the time-domain search and flags about 150 to 250 channels (about 10\% of the band) at any time. A third RFI search is applied to the final maps using another median filter algorithm. It is usual for an additional 10\% of the band to be flagged during this stage, which leaves about 80\% of the band usable for radio astronomy.

Fortunately, the Observatory site is in mountainous country and terrain shielding very effectively attenuates interfering signals from terrestrial sources. Interference from satellites and airborne transmitters is not overwhelming. With the implementation of the three relatively simple RFI excision procedures it has been possible to make sensitive observations over a band in which radio astronomy has little protection from other spectrum users. 

\section{Stability \& Polarization Performance}
\label{StabilityAndPolPerf}

\begin{figure}[tbp]
  \includegraphics[clip, width=\columnwidth, bb=34 50 710 525]{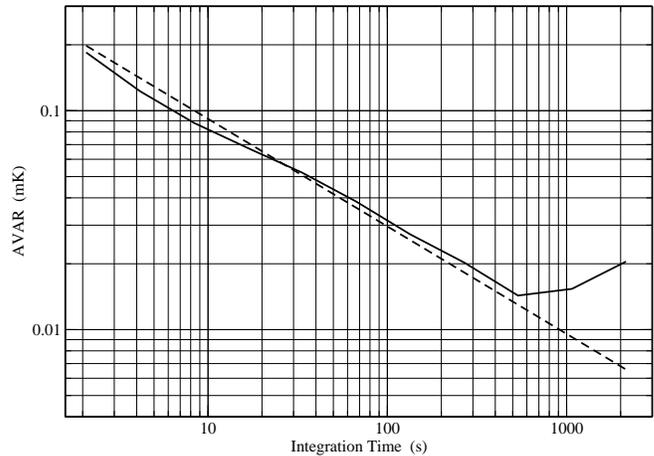}
   \caption{Allan-variance plot of total power data in a single
   channel at 1460~MHz, measured with the telescope pointing towards
   the North celestial pole (continuous line). For reference the
   Allan variance of white noise is plotted (dashed line). The receiver is stable over periods of at least
   500s. Longer term drifts are probably caused by changes in the level of
   Galactic emission received through the sidelobes during the measurement. }
\label{figure5}
\end{figure}

The short-term stability of the whole system was evaluated by an Allan-Variance measurement with the telescope pointing towards the North celestial pole. The Allan variance plot (Fig.~\ref{figure5}) shows that received noise behaves like white noise and no systematic drifts occur on time scales of less than 500 seconds. Variations on timescales longer than 500 seconds probably arise from changing signal levels as the sky drifts through the telescope sidelobes.

The long-term stability of the system was examined based on the daily calibration sources. A variation in gain of the whole system of 1\% over periods of one month was found, possibly caused by small temperature variations inside the focus box. These small instabilities are removed by calibration against astronomical sources.

Instrumental polarization has two components, cross-talk between R and L channels (occuring in the feed and waveguide components and their interaction with reflections at the low-noise amplifier inputs). Cross-talk varies from 25\% at the bottom end of the band to $\lesssim$ 3\% towards the high frequency end. The cross-polar response of the antenna has not been taken into account in the results presented here; this effect is evident only where total intensity is very high  - along the Galactic plane.

\section{Proof-of-Concept Study}

The first 8\% of data (300 of a total of 3600 scans) of the GMIMS component survey with the DRAO 26-m Telescope are used here for a proof-of-concept study. The goal of this study is to develop processing, calibration, and analysis algorithms, which will eventually be applied to the complete DRAO survey as well as to the other component surveys carried out by the GMIMS collaboration. The result of this study is the first RM-Synthesis of Galactic, diffuse polarized emission, obtained with a single-antenna telescope.

\subsection{Observations}

A set of scans along the Meridian, 12 arcmin spaced, was defined which would provide sampling slighly surpassing the Nyquist criterion of the entire sky from  $-30^{\circ}$ to $87^{\circ}$ declination at the highest frequency. Long scans extend from the horizon limit at $-30^{\circ}$ declination to $87^{\circ}$, close to the North celestial pole, and short scans from $-30^{\circ}$ to $60^{\circ}$. To avoid oversampling around the pole every second scan is a short scan. A quasi-random sequence of scans was devised, aiming to avoid systematic effects while making optimal use of the observing time. With a scanning speed of 52.5 arcmin/min short scans take 1.7 hours and long scans 2.2 hours. Scans are made exclusively at night to avoid contributions from solar emission through the sidelobes, but calibration sources are observed during the hour before sunset and after sunrise. For the final survey each scan will be observed twice, as an up and a down scan, resulting in double coverage of every pixel. The region presented in this paper was observed in April 2008 and is non-uniformly covered.

Resolution was reduced to $1^\circ$ by smoothing in the image plane to approximate full Nyquist sampling. The rms noise in one channel (237~kHz width) was found to be $\sim 80$ mK in total power and $\sim 40$ mK in polarization. This is about 1.6 times higher than the theoretical value of 25 mK, calculated based on our system temperature, bandwidth, and integration time. The reasons for the higher rms noise are a) the deactivation of one of the four FFT lanes on each FPGA due to temperature problems\footnote{The problem has since been fixed and all four lanes are now in operation.}, which resulted in a loss of 25\% of digitized samples and thus increased the rms noise by a factor of 1.15; and b) weighting of the digitized samples using a Blackman-Harris windows function, which, in effect, made use of only 50\% of the input signal and thus increased rms noise by a factor of 1.41.

\subsection{Post-Processing \& Calibration}

We now discuss the most important post-processing and calibration steps. These are: 1) correction of instrumental effects (including corrections for bandpass, relative phase shift across the band, and cross-talk introduced in the receiving system by a left polarized signal entering the right channel and vice-versa); 2) subtraction of ground radiation; and 3) removal of scanning effects. A correction of the cross-polar response of the telescope beam, which requires a deconvolution of the final maps with the beam pattern, was not attempted. While bandpass and phase corrections were averaged over a few channels, all other corrections were performed on each channel individually. 

\begin{figure}[tbp]
  \includegraphics[clip, width=\columnwidth, bb=34 50 754 528]{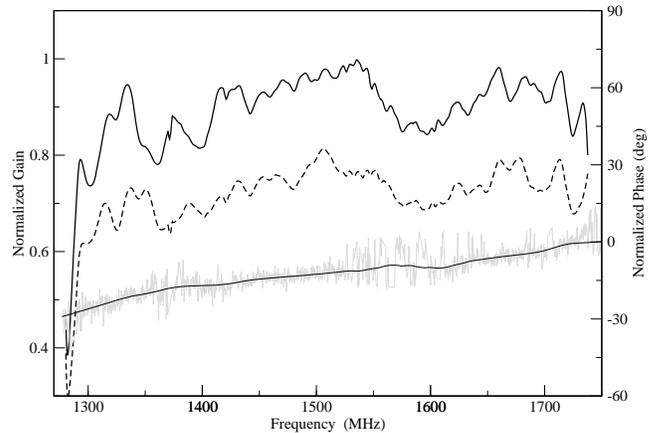}
  \caption{Relative gain of RR (continuous) and LL (dashed) on the left axis, as obtained through a single observation of Tau-A. The right axis shows the phase difference (RR $-$ LL) across the band as measured through a series of 3C286 and 3C270 observations. Plotted are the actual measurement (gray) and a smoothed version (thin black) that is used for the calibration. The phase is set to zero at the highest frequency.}
\label{figure6}
\end{figure}

In addition to the electronic gain calibration, which is part of the real-time processing chain (Sect.~\ref{egainphase}), daily flux calibrations are derived from observations of two of our four flux calibrators (see Fig.~\ref{figure6}). We used Cyg-A, Tau-A, and Vir-A as our primary calibrators. Cas-A was used as secondary calibrator whenever none of the primary calibrators was visible. The flux of the secondary calibrator was derived from the primary calibrators. A calibration observation consisted of 11 declination scans spaced by 12 arcmin, yielding a map of the source. A two-dimensional Gaussian was fitted to the calibration observations and the amplitude of the fitted Gaussian was used as the calibration value. Table~\ref{table2} summarizes the flux values and spectral indices ($\alpha$) used for the primary calibrator and derived for the secondary calibrators. We found that flux values slightly different from the published values gave the best overall consistency across the four calibrators. The final data are tentatively calibrated in main-beam brightness temperature by assuming an aperture efficiency of 53\% over the entire band, a value originally determined for 1.4 GHz by \citet{2000AJ....120.2471H}. In the future, when survey data over a much larger period need to be calibrated, it is planned to take intrinsic variability of Cas-A into account. 

\begin{deluxetable}{llllll}[tbhp]
\tablecaption{Primary and secondary gain calibrators}
\tablehead{
\colhead{Name} & \multicolumn{2}{c}{Literature (1.4 GHz)} & \multicolumn{2}{c}{Adopted (1.4 GHz)} & \colhead{Remarks} \\
\colhead{} & \colhead{Flux} & \colhead{$\alpha$} & \colhead{Flux} & \colhead{$\alpha$} & \colhead{}} \\
\startdata
Cyg-A  	& 1579 Jy &  -1.01 	& 1589 Jy &  -1.07 & 1,3\\
Tau-A 	& 908 Jy & -0.28 	& 848 Jy & -0.27 & 1,3\\
Vir-A 	& 208 Jy & -0.83	& 207 Jy & -0.90 & 1,3 \\
Cas-A 	& 2442 Jy & -0.78	& 1861 Jy & -0.77 & 2,4
\enddata
\footnotetext[1]{Primary Calibrator.}
\footnotetext[2]{Secondary Calibrator.}
\footnotetext[3]{Flux value taken from the VLSS Bright Source Spectral Calculator (Helmboldt et al. 2008), which is based on Baars et al. (1977).}
\footnotetext[4]{The literature flux value is the value for the epoch 1980 from Baars et al. (1977).}
\label{table2}
\end{deluxetable}

The relative phase between the two polarization channels (R and L) was unknown and was determined through observations of 3C286 and 3C270. Both sources are at high Galactic latitudes and have well known RMs of $\approx 0$. At 1.4 GHz, the percentage polarization and polarization angles of 3C286 and 3C270 are 9.3\% and $32^\circ$, and 8.0\% and $123^\circ$, respectively \citep{1980A&AS...39..379T}. This measurement revealed a phase gradient of about 30\degr~across the band (Fig.~\ref{figure6}). A phase calibration is not done on a regular basis since this gradient will be stable until changes to the receiver are made that affect the path length or delay of the signal paths.

Cross-coupling of one hand of polarization into the other results in instrumental polarization. In the final polarization maps this effect leads to a conversion of Stokes~I into U and Q, leading to the additional signals $\mathrm{U}_{\mathrm{inst}} = f_\mathrm{U}\,\mathrm{I}$ and $\mathrm{Q}_{\mathrm{inst}} = f_\mathrm{Q}\,\mathrm{I}$. This type of instrumental polarization is the result of imperfections in the feed and mismatches in the receiver.  The coefficients $f_\mathrm{U}$ and $f_\mathrm{Q}$ were determined daily from the flux calibration. The integrated polarization from the primary calibrators is very close to zero and they were considered to be unpolarized. Any polarization detected is then of instrumental origin. As it is evident in the final maps, the removal of cross-talk is not sufficient for a complete correction for instrumental polarization. The images still show spurious polarization features around bright sources produced by the cross-polar response of the telescope. The correction of cross-polar response will require a cleaning (de-convolution) of the final polarization data, which is planned for the complete survey and requires fully Nyquist-sampled maps in all Stokes parameters.

Ground radiation is received through the side- and spillover lobes of a telescope. It can contribute a signal an order of magnitude stronger than the sky, especially at low elevations. As a temporary approach for the data presented in this paper, ground radiation profiles were determined by plotting the observed Stokes I, U, and Q intensities versus elevation. Only scans made between 13h and 17h R.A. were used, avoiding the bright emission from the Galactic plane. Ground contribution was then retrieved from these scatter plots by taking the lower envelope in Stokes I, and a median value for Stokes U and Q. 

\begin{figure}[tbp]
\includegraphics[clip,width=\columnwidth,bb= 45 305 630 714]{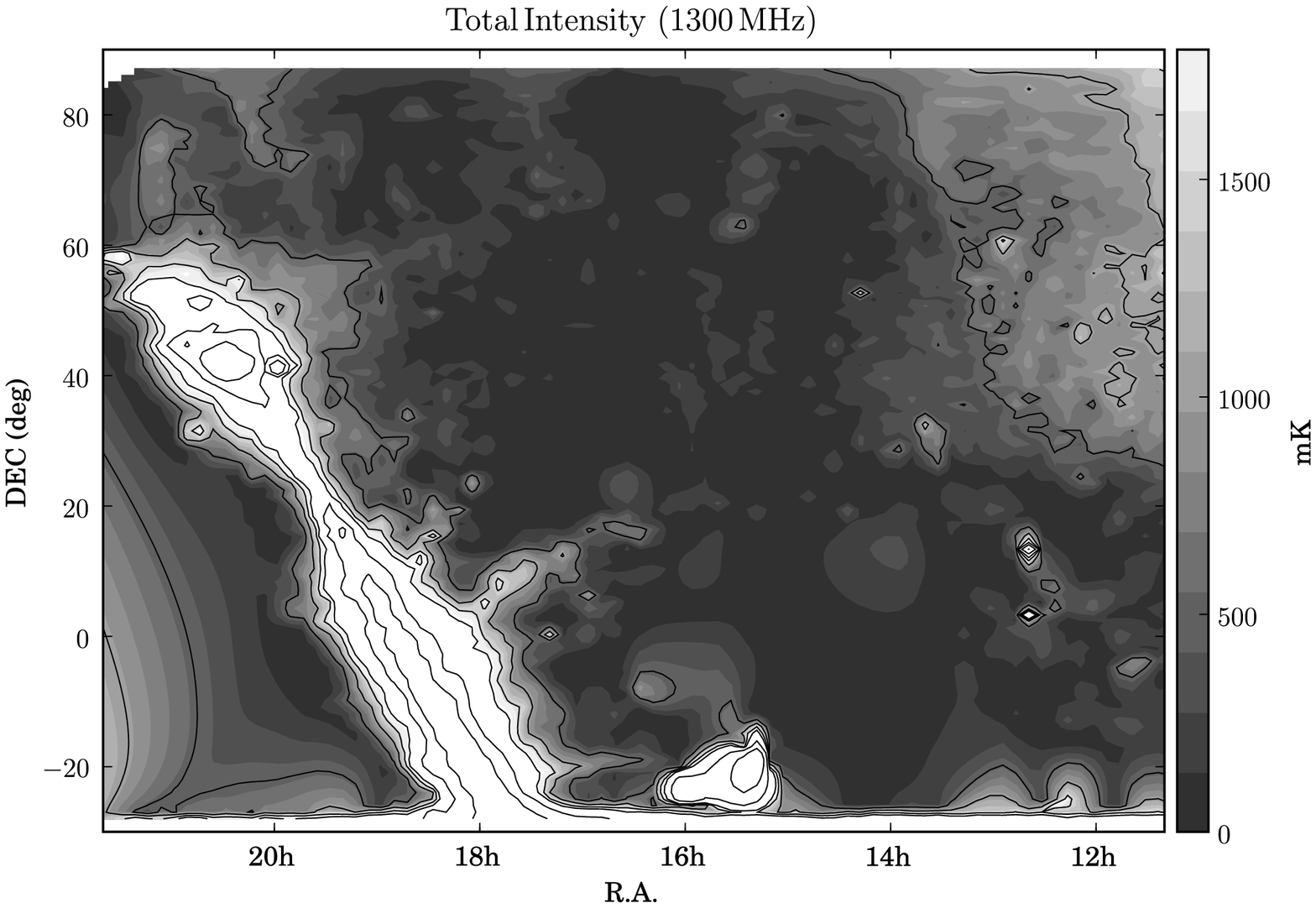}
\includegraphics[clip,width=\columnwidth,bb= 45 305 630 714]{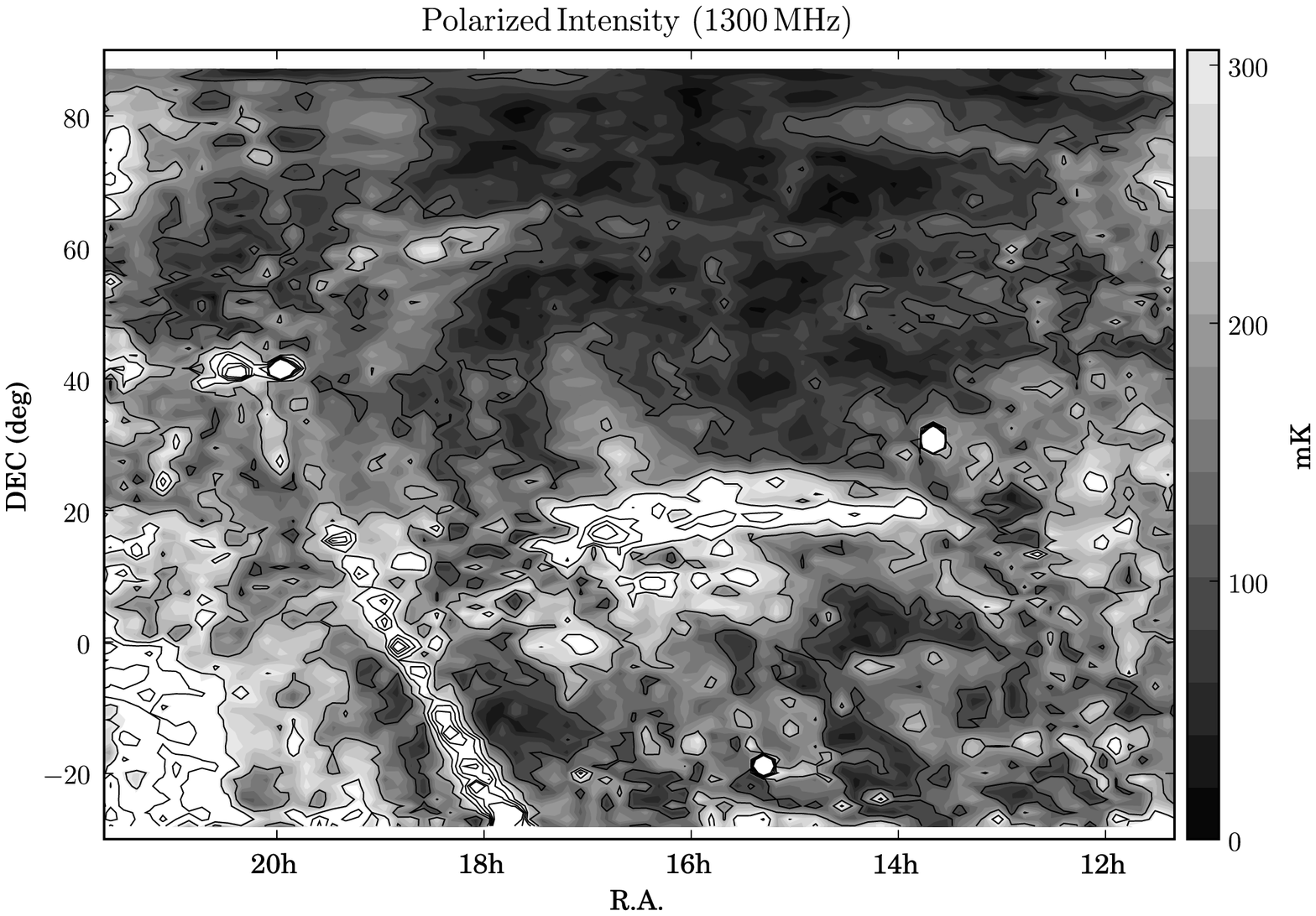}
\includegraphics[clip,width=\columnwidth,bb= 45 305 630 714]{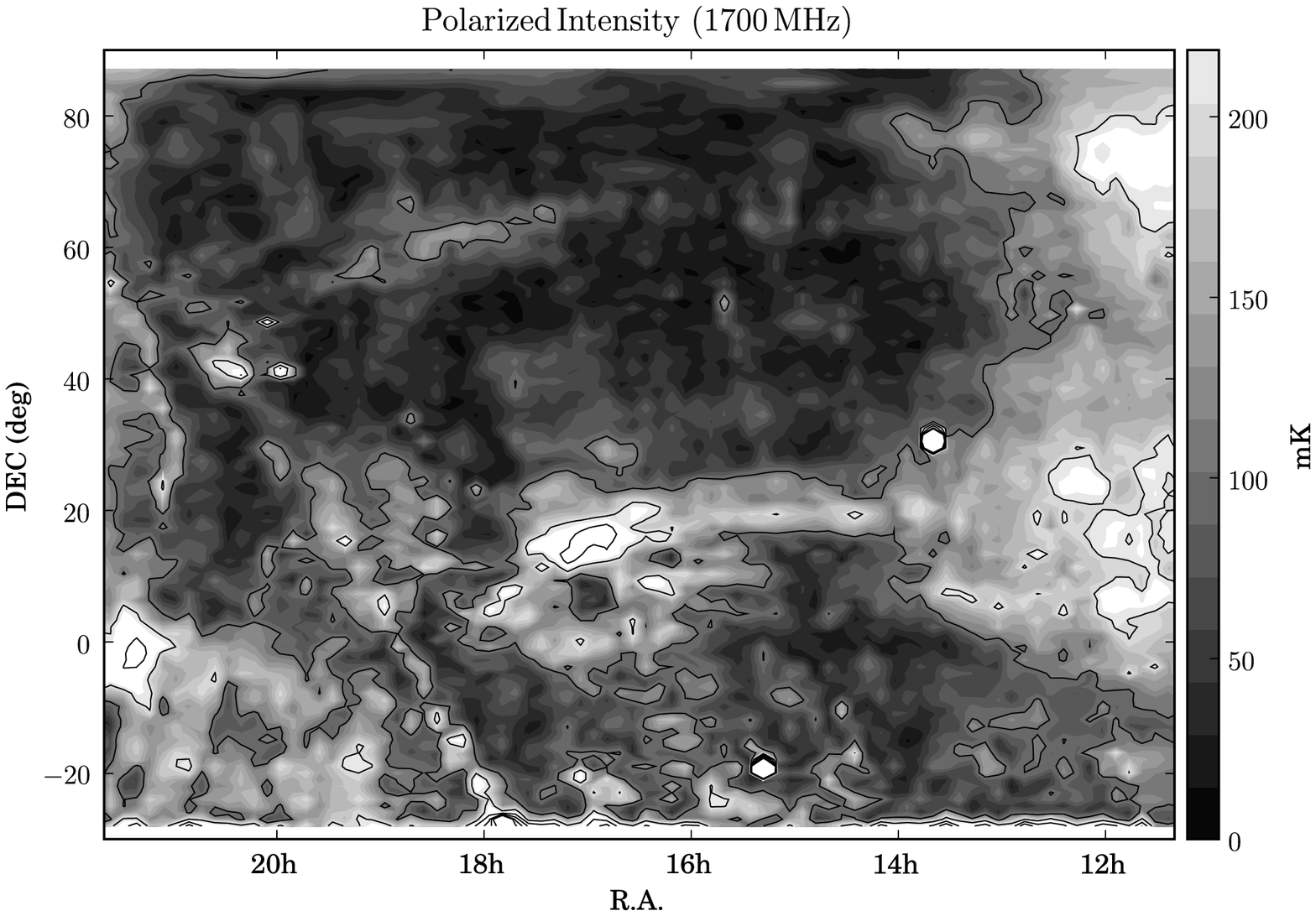}
\caption{Maps show the first 8\% of survey data, tentatively calibrated as explained in the text. Total intensity is shown at 1300 MHz, and polarized intensity is shown at 1300 and 1700 MHz. Contour lines in total intensity correspond to 500, 1000, 1500, 2000, 4000, and 8000 mK. In polarized intensity contour lines indicate polarized intensity in steps of 100 mK at 1300 MHz, and 50 mK at 1700 MHz, resp. Features in the maps are explained in the text.}
\label{figure7}
\end{figure}

Scanning effects (stripes) in the final maps were removed by a technique known as ``basket weaving''. This technique removes baseline effects by comparing the overlaps of scans made in different directions and times with the assumtion that the baselines vary and the sky brightness remains constant \citep{1979A&A....74..361S}. The stripes are most likely caused by variations of the electromagnetic properties of the ground surrounding the telescope affecting the ground pick-up, time-variability of the noise figure of the LNAs,  or small errors in the flux calibration. Basket weaving was applied on each channel individually. To cope with the high data volume the basket weaving algorithm ran on a cluster consisting of 40 CPUs.  A problem with the basket weaving technique is the floating baseline level which can introduce large-scale distortions in the maps. The Stokes I map is more affected by these distortions then the Stokes U and Q maps. The maps presented here (Fig.~\ref{figure7}) show the R.A. range mostly free of distortions (a gradient is still present in the total intensity map at R.A. $> 20$h).

\subsection{Application of Rotation Measure Synthesis}

RM-Synthesis consists of two separate steps: 1) de-rotation of Stokes U and Q, and 2) cleaning of the RM-Spectra. In the first step, the final data cube is de-rotated for a range of RM values. Here we used RMs from -1250 to 1250 \radm2 in steps of 10 \radm2. For each RM value the Stokes U and Q values in each pixel and channel are de-rotated by the  angle $\mathrm{RM}\,\lambda^2$. The de-rotated cube is then collapsed (averaged) into a 2-D map of polarized intensities, which forms a slice of the RM-Synthesis cube. RM-Spectra are constructed for each pixel from the series of 2-D maps at different RMs. 

\begin{figure}[tbp]
  \includegraphics[clip, bb=41 44 705 525, width=\columnwidth]{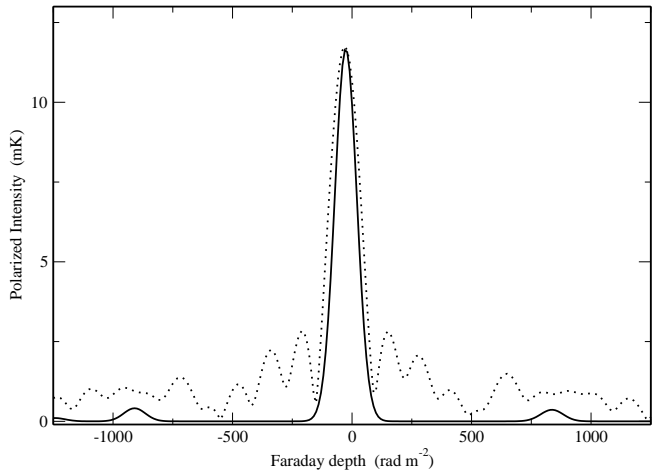}
  \caption{A cleaned (solid line) and uncleaned (dotted line) RM-Spectrum for a position close to the centre of the surveyed region. The uncleaned spectrum represents a typical RMSF for our data.}
\label{figure9}
\end{figure}

The second part of RM-Synthesis is cleaning. Cleaning is necessary because the observed spectra are ``contaminated'' with spurious line features caused by sidelobes in the rotation measure spread function (RMSF). The RMSF is the response function (the instrumental response) with which the spectrum is convolved and is the result of incomplete frequency coverage. Figure~\ref{figure9} shows an RMSF for our observations, which is a typical RMSF for data in our frequency range, coverage, and flagging. Using an analogy to synthesis imaging, the RMSF can be understood as the ``dirty beam''. The observed spectra (``dirty images'') need to be de-convolved (with an appropriate clean algorithm) to obtain the ``cleaned images''. The clean model consists of a series of delta peaks. To obtain cleaned RM-spectra the RM-model is convolved with a Gaussian (the ``restoring beam'') with a width corresponding to the expected resolution in RM-Synthesis, which, in our case, was slightly less than 132 \radm2. Figure~\ref{figure9} shows a cleaned and an uncleaned RM-Spectrum. Spurious components at the $\sim 5$\% level are still present and are visible at $\pm 850$ \radm2 in the spectrum shown.

Our frequency range and spectral resolution determine: 1) the resolution of the RM-Spectra (the resolution in Faraday depth), 2) the largest scale in Faraday depth we are sensitive to, and 3) the maximum observable Faraday depth. These values are summarized in Tab.~\ref{table3}. In our case the largest scale in Faraday depth we can measure is smaller than the resolution. This is an important constraint to keep in mind: two peaks in our RM-spectrum do not necessarily indicate two different Faraday rotation layers, but may also arise from a continuous distribution of RMs in Faraday depth, a few RMTFs wide, of which only the two edges are resolved.

\begin{deluxetable}{ll}[tbhp]
\tablecaption{RM-Synthesis Parameters}
\tablehead{
\colhead{Parameter} & \colhead{Value} } \\
\startdata
Resolution in Faraday Depth & 132~\radm2 \\
Largest Scale in Faraday Depth & 108~\radm2 \\
Maximum Observable Faraday Depth & 10$^5$~\radm2 
\enddata
\label{table3}
\end{deluxetable}

\subsection{General Discussion}

\begin{figure}[tbp]
\includegraphics[clip,width=\columnwidth,bb= 45 305 623 714]{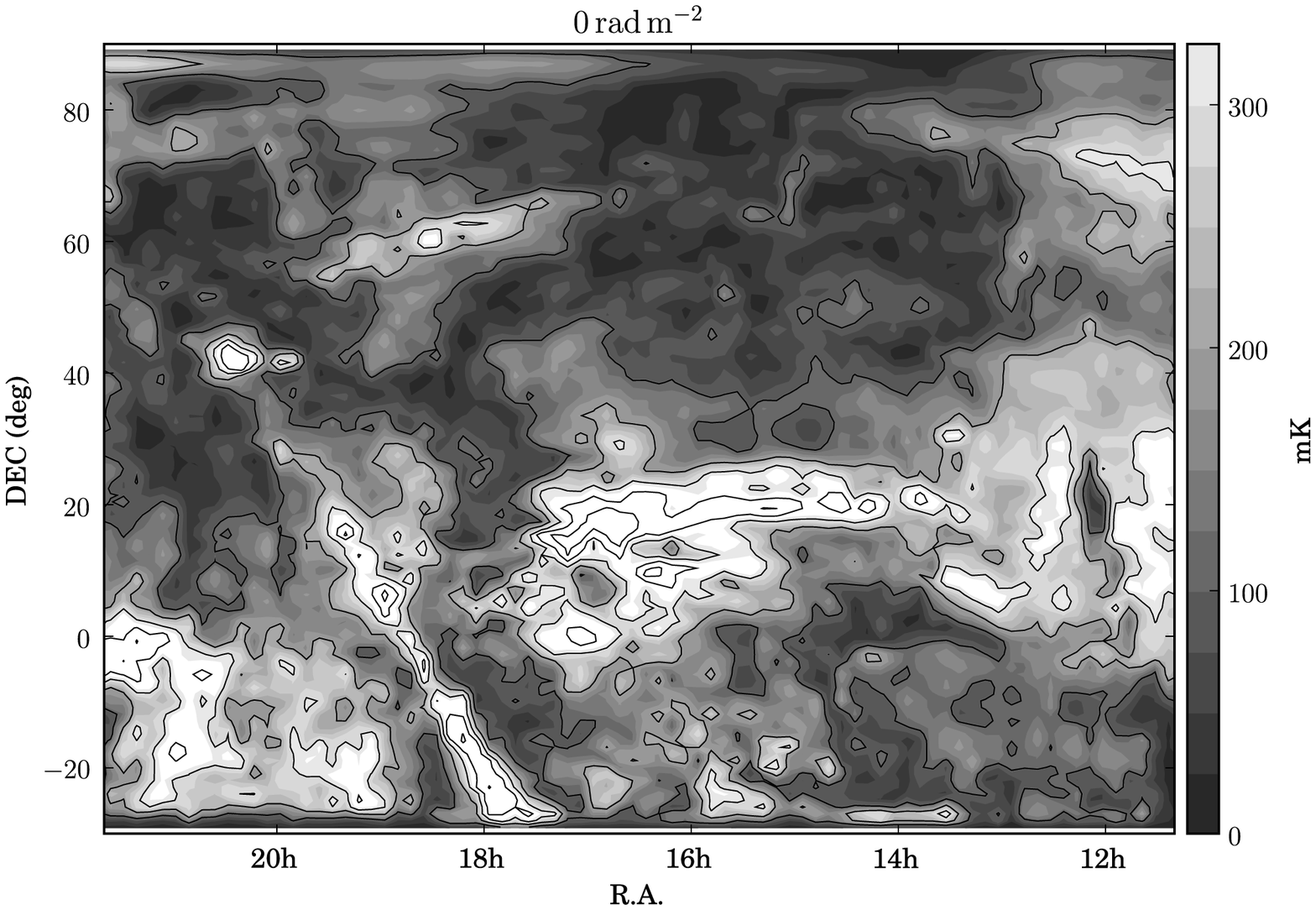}
\includegraphics[clip,width=\columnwidth,bb= 45 305 623 714]{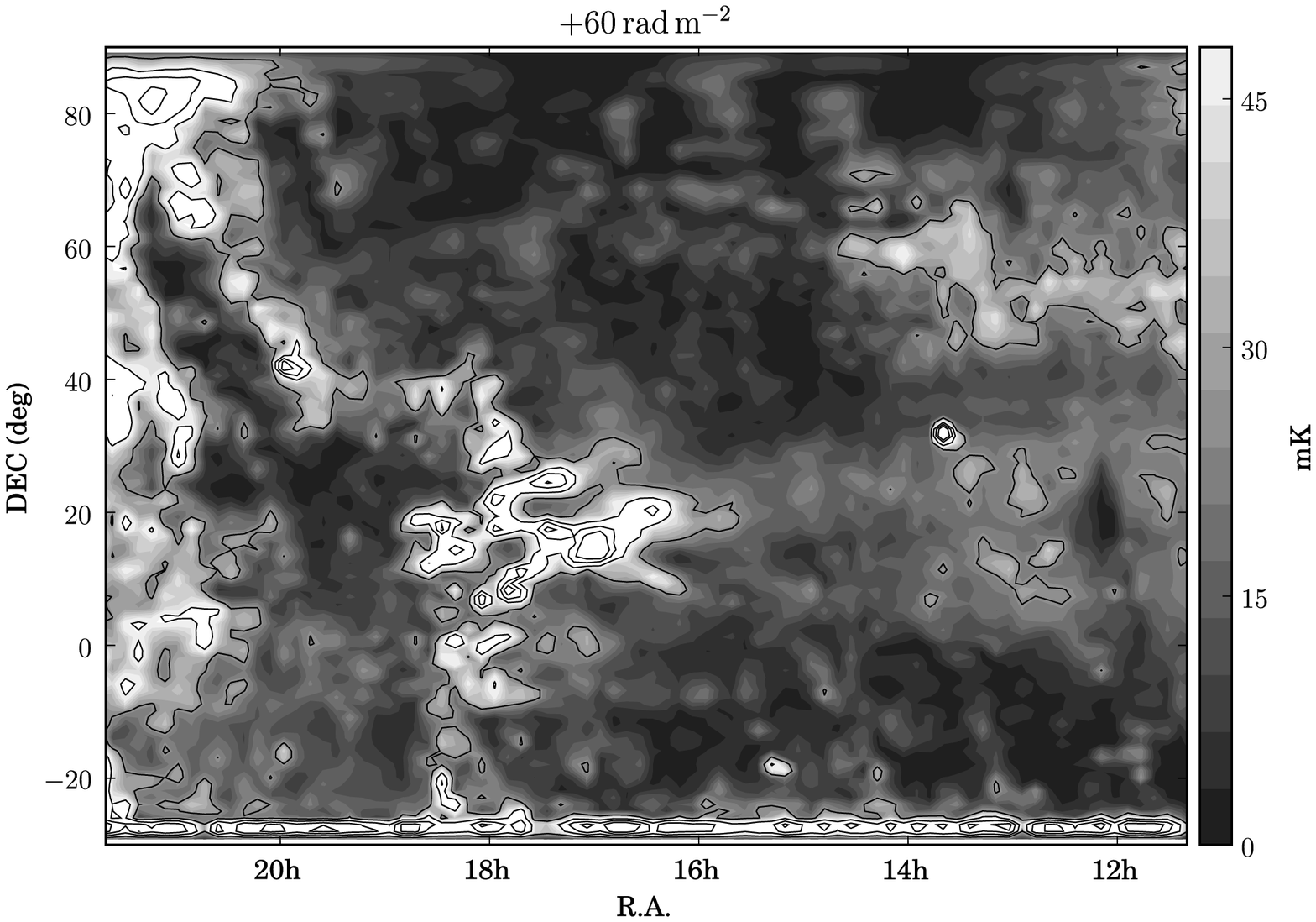}
\caption{The upper map shows the RM-Synthesis slice at 0 \radm2 with contour lines  every 100 mK of polarized intensity. The lower map shows the 60 \radm2 slice with contour lines in steps of 25 mK.} 
\label{figure10}
\end{figure}

Figure~\ref{figure10} shows two slices of the RM-Synthesis cube. The rms sensitivity of these images is $\sim 3$~mK, the rms noise expected from the entire 470~MHz bandwidth. This high sensitivity is one of  the great benefits of RM-Synthesis. Furthermore, the abundant frequency coverage permits quite aggressive flagging of RFI, leading to good signal-to-noise ratio even outside the protected radio astronomy bands. Finally, we note that RM-Synthesis eliminates the problem of bandwidth depolarization: it provides an adaptive filter that can be tuned to the characteristics of the emission.

The strength of RM-synthesis, however, is in revealing 3-D structures in Faraday space. This requires a $\lambda^2$-coverage matched to the expected Faraday rotation. Our resolution in Faraday depth is probably sufficient to map the general trend of RMs, but will only allow us to resolve out multiple RM components for highly Faraday rotated parts of the ISM. Our frequency range ($\lambda^2$-coverage), however, will allow us to detect highly Faraday rotated structures even if the fractional polarization of these regions is extremely low. Without RM-Synthesis the data would be ``confusion limited'' in the sense that faint but highly Faraday rotated emission could not be separated from the usually stronger, unrotated ``foreground''. RM-Synthesis helps by separating differently Faraday rotated structures.

The data presented here are preliminary products in the sense that they are not yet absolutely calibrated: ground radiation profiles were determined in an area of the sky containing large-scale Galactic emission in Stokes I, U, and Q, which makes the separation of ground and sky emission difficult. An independent, absolute temperature scale has not been established for the telescope. Instrumental polarization is visible in the maps caused by the cross-polar response of the telescope, which is most obvious along the Galactic plane. Smaller problems are remaining scanning effects. Despite the preliminary state of these data, the final Stokes U and Q maps at 1.4 GHz compare very well with the DRAO Low Resolution Polarization Survey at 1.4 GHz \citep{2006A&A...448..411W}, which convinces us that our proof-of-concept study will provide valuable input for future RM-Synthesis observations with single-antenna telescopes. 

\subsection{Polarized Emission at 0 \radm2}

A few large-scale structures in the 0 \radm2 RM-Synthesis slice (Fig.~\ref{figure10} top) are immediately obvious. Polarized emission from the North-Polar Spur is clearly visible as an arc-like filament in the centre of the map, going  in R.A. from 14h to 17h30m, and in DEC from $-5^\circ$ to $20^\circ$. South of the Galactic plane, at R.A. from about 19h to 21h30m and DEC from  $-25^\circ$ to $0^\circ$, a patch of polarized emission is visible, which has been interpreted as part of the Loop~I structure \citep{2007ApJ...664..349W}. At R.A. = 18h30m and DEC = $60^\circ$  another polarized filament is visible, most likely associated with Loop~III. Polarized emission within a narrow strip along the Galactic plane can be assumed to be instrumental polarization.

\subsection{Polarized Emission at High Positive RMs}

The RM-Synthesis slice at 60 \radm2 (Fig.~\ref{figure10} bottom) is chosen as an example for polarized emission at high positive RMs. Most features in this slice actually peak at slightly different RMs. The most obvious structure is centred on R.A. = 17h30m and DEC$=20$\degr, about $20$\degr~in size. Other structures along the Galactic plane and structures close to the edges of the map may be spurious. 

Previous polarization surveys could detect polarized emission from the North-Polar Spur only at Galactic latitudes above $\sim 25^\circ$, while in total intensity the spur can clearly be traced down to lower latitudes. The two filaments found at high positive RMs run from the southern tip of the polarized emission of the North-Polar Spur to the Galactic plane. The two filaments may therefore be associated with the North-Polar Spur, perhaps representing highly Faraday-rotated low-latitude emission of the spur. We postpone further analysis and interpretation until sampling and angular resolution are improved.

In summary, the two filaments at 60 \radm2 are good examples of structures that are only visible in polarization and which can be revealed only by RM-Synthesis. Such high RM require a strong magnetic field parallel to the line-of-sight. The filaments are most likely structures in the magnetic field, perhaps produced by supernovae or stellar winds in the interstellar medium, compressing the ambient magnetic field.

\subsection{Polarized Emission at Negative RMs}

Our RM-Synthesis data reveal faint, Faraday-rotated structures  at RM $\lesssim -25$ \radm2. Some of these structures are filamentary in shape and seem to be associated with the North-Polar Spur, although not with its southern tip but other parts of its shell at higher Galactic latitudes. Thus, it is possible that our RM-Synthesis data reveal, in fact, the 3-D structure of the magnetic field of the North-Polar Spur, but more complete coverage of this region is required to confirm this interpretation. 

\subsection{Conclusions from the Proof-of-Concept Study}

Wide-band polarimetric data present a challenge for calibration in that the receiver characteristics and performance can vary significantly over the band. Furthermore, the orientation and magnitude of sidelobes and spillover lobes depends on frequency, resulting in highly frequency-dependent ground radiation profiles. In our data we find channels that are virtually free of instrumental polarization while other channels are almost free of ground radiation.

These effects can be very useful for calibration. For example, channels free of ground radiation can be used as ``anchor points'' for an absolute calibration, making the removal of ground radiation more robust in the presence of large-scale emission structures in the data. Similarly, channels free (or nearly so) of cross-talk can be very useful for the determination of the cross-polar component of the instrumental polarization.

An important observation is that we find unexpectedly high RM values for the diffuse emission. Diffuse emission is subject to various depolarization effects along the line-of-sight, which has led to the argument that most of the diffuse polarized emission at $\lesssim 2$~GHz must be of local origin within a few kpc. We find RMs as high as 100~\radm2 for some of the diffuse emission, which may either indicate that the emission comes from a larger distance, or that the local magnetic field is stronger than previously believed.

Previous rotation measure surveys of the diffuse Galactic emission revealed only moderate RMs of less than a few tens of \radm2 \citep{1976A&AS...26..129B}. We conclude that Faraday rotation should be measured using wide-band spectro-polarimetric observations. The relation between polarization angle and $\lambda^2$ is not always linear for the diffuse Galactic emission, rendering the traditional techniques of determining RM by least-square fitting inadequate for diffuse emission.

\section{Summary \& Outlook}

Recent developments in digital signal processing allowed us to build a digital polarimeter in a short time. In this paper we described the technical process from building a digital polarimeter, designing the feed, and obtaining, calibrating, and analyzing wide-band polarization data. For the first time, RM-Synthesis was successfully applied to wide-band spectro-polarimetric observations of the diffuse Galactic emission obtained with a single-antenna telescope over a large region on the sky. 

Observations for the DRAO Rotation Measure Survey are ongoing and are 42\% completed as of September 2009. The final survey will be better than Nyquist sampled with an rms sensitivity, in a 12 arcmin wide pixel, of 105 mK per channel or 3 mK for the entire band (taking into account data loss due to RFI). Although not yet in a publishable state, a first look at these data confirms the presence of the polarized structures discussed in this paper. 

This survey forms part of GMIMS. Three of the six component surveys of the GMIMS project are now underway, with another two surveys in the planning stages. An important objective of GMIMS is that all measurements be absolutely
calibrated. In practical terms this means that the calibration noise
signal will  be carefully measured relative to terminations at cryogenic
temperatures (liquid nitrogen), and that the aperture efficiency will be
determined using calibrators of well-known flux density. 

\acknowledgements{}

The authors would like to thank Sundance and 3L for outstanding support during the development phase of the digital polarimeter. We are grateful to the IT staff at DRAO, Tony Hoffman and Peter Cimbaro, for providing and maintaining the telescope infrastructure. We also thank Ron Casorso, Paul Dunlop, and Dale Basnett for rewiring and recabling the telescope.  The Dominion Radio Astrophysical Observatory is a National Facility operated by
the National Research Council Canada.

\end{document}